\pdfoutput=1
\documentclass[10pt]{article}
\usepackage[T1]{fontenc}
\usepackage[english]{babel}
\usepackage{amsmath}
\usepackage[backend=bibtex,
  style=ieee,
  sorting=none,
  url=false,
  style=numeric,
  maxbibnames=99,
firstinits=true]{biblatex}
\renewbibmacro{in:}{}
\bibliography{/home/rsrsl/library/jabref_biblio.bib}
\usepackage{csquotes}
\bibliography{~/library/jabref_biblio.bib}
\usepackage{booktabs}
\usepackage{amsthm}
\usepackage{lmodern}
\usepackage{amsfonts}
\usepackage{amssymb}
\usepackage{graphicx}
\usepackage[kerning=true,tracking=true,final]{microtype}
\usepackage{color}
\usepackage{algorithm}

\newcommand{\Tr} {\mbox{\rm tr}}

\newcommand{\diag} {\mbox{\rm diag}}

\newcommand{\E}{\mathbb{E}}

\newcommand{\R}{\mathbf{R}}
\newcommand{\kron}{\otimes}
\newcommand{\N}{\ensuremath{\mathcal{N}}}
\renewcommand{\matrix}[1]{\begin{bmatrix} #1 \end{bmatrix}}

\newtheorem{theorem}{Theorem}

\newtheorem{assumption}{Assumption}

\newtheorem{remark}{Remark}

\title{\LARGE \bf On the estimation of initial conditions\\in kernel-based system identification}
\author{Riccardo S. Risuleo, Giulio Bottegal and H\r akan Hjalmarsson
\thanks{R. S. Risuleo, G. Bottegal and H. Hjalmarsson are with the ACCESS Linnaeus Center, School of Electrical Engineering, KTH Royal Institute of Technology, Sweden
        (e-mail addresses: {\tt\small risuleo@kth.se}\,, {\tt\small bottegal@kth.se}\,,
        {\tt\small hjalmars@kth.se}). This work was supported by the European
    Research Council under the advanced grant LEARN, contract 267381 and by the
Swedish Research Council under contract 621-2009-4017.}} %chktex 8
\begin{document}
\maketitle

\begin{abstract}
Recent developments in system identification have brought attention to
regularized kernel-based methods, where, adopting the recently introduced stable
spline kernel, prior information on the unknown process is enforced. This
reduces the variance of the estimates and thus makes kernel-based methods
particularly attractive when few input-output data samples are available. In
such cases however, the influence of the system initial conditions may have a
significant impact on the output dynamics. In this paper, we specifically
address this point. We propose three methods that deal with the estimation of
initial conditions using different types of information. The methods consist in
various mixed maximum likelihood--a posteriori estimators which %chktex 8
estimate the  initial conditions and tune the hyperparameters characterizing the
stable spline kernel.  To solve the related optimization problems, we resort to
the expectation-maximization method, showing that the solutions can be attained
by iterating among simple update steps. Numerical experiments show the
advantages, in terms of accuracy in reconstructing the system impulse response,
of the proposed strategies, compared to other kernel-based schemes not
accounting for the effect initial conditions.
\end{abstract}

\section{Introduction}
Regularized regression has a long history~\cite{tikhonov1977solutions}. It has
become a standard tool in applied statistics~\cite{hastie2005elements}, mainly
due to its capability of reducing the mean square error (MSE) of the regressor
estimate~\cite{james1961estimation}, when compared to standard least
squares~\cite{ljung1999system}.  Recently, a novel method based on regularization has been
proposed for system identification~\cite{pillonetto2014kernel}. In this
approach, the goal is to get an estimate of the impulse response of the system,
using the so called kernel-based methods~\cite{scholkopf2002learning}. To this end, the
class of \emph{stable spline kernels} has been proposed recently
in~\cite{pillonetto2010new},~\cite{pillonetto2011prediction}. The main feature of these kernels is that they
encode prior information on the exponential stability of the system and on the
smoothness of the impulse response. These features have made stable spline
kernels suitable for other estimation problems, such as the reconstruction of
exponential decays~\cite{pillonetto2010regularized} and correlation functions~\cite{bottegal2013regularized}. Other kernels for system identification have
been introduced in subsequent studies, see for instance~\cite{chen2012estimation},~\cite{chen2014constructive}.

Stable spline kernels are parameterized by two \emph{hyperparameters}, that
determine magnitude and shape of the kernel and that need to be estimated from
data. An effective approach for hyperparameter estimation relies upon \emph{
empirical Bayes} arguments~\cite{maritz1989empirical}. Specifically, exploiting the
Bayesian interpretation of regularization~\cite{wahba1990spline}, the impulse response
is modeled as the realization of a Gaussian process whose covariance matrix
corresponds to the kernel. The hyperparameters are then estimated by maximizing
the marginal likelihood of the output data, obtained by integrating out the
dependence on the impulse response. Given a choice of hyperparameters, the
unknown impulse response is found by computing its minimum MSE Bayesian
estimate~\cite{pillonetto2014kernel}.

One situation where kernel-based methods are preferable is when data
records are short (e.g., five times the rise time of the system). This mainly
because of two reasons:
\begin{enumerate}
\item Kernel-based methods do not require the selection of a model order.
  Standard parametric techniques (such as the prediction error
  method~\cite{ljung1999system},~\cite{soederstroem1988system}) need to rely on model selection criteria,
  such as AIC or BIC, if the structure of the system is
  unknown~\cite{beghelli1990frisch}. These could be unreliable when
  faced with small data sets.
\item The bias introduced by regularization reduces the variance. With small
  data records, the variance can be very high. If the bias is of the right kind,
  it will compensate for the variance effect in the MSE~\cite[Ch.
  2.9]{hastie2005elements}.
\end{enumerate}

When data records are very short (e.g., two times the rise time of the system) we
cannot ignore the effect of the initial conditions. In fact, if the system is
not at rest before the experiment is performed, then there are transient effects
that cannot be explained using only the collected data. Standard workarounds,
such as discarding those output samples that depend on the initial conditions or
approximating the initial conditions to zero~\cite[Ch. 10.1]{ljung1999system}, may give
unsatisfactory results. Thus, it seems preferable to deal with the initial
conditions by estimating them. In this paper we discuss how to incorporate the
estimation of the initial conditions in the context of kernel-based system
identification. We discuss three possible approaches to the problem. First, we
propose a method that incorporates the unknown
initial conditions as parameters, to be estimated along with the kernel
hyperparameters.  Then, assuming that the input is an
autoregressive--moving-average (ARMA) stationary process, we propose %chktex 8
to estimate the
initial conditions using the available samples of the input, thus designing a
minimum variance estimate of the initial conditions from the input
samples. Finally, we design a mixed maximum a posteriori--marginal %chktex 8
likelihood (MAP--ML) estimator (see~\cite{yeredor2000joint}) that %chktex 8
effectively exploits information from both input and output data. We solve the
optimization problems using novel iterative schemes based on the
expectation-maximization (EM) method~\cite{dempster1977maximum}, similar to the
technique used in our previous works~\cite{risuleo2015kernel}
and~\cite{bottegal2015blind}, where methods for Hammerstein and blind system
identification are proposed. We show that each iteration consists of a set of
simple update rules which either are available in closed-form or involve scalar
optimization problems, that can be solved using a computationally efficient grid
search.

The paper is organized as follows. In Section~\ref{sec:Problem}, we formulate
the problem of system identification with uncertainty on the initial conditions.
In Section~\ref{sec:Bayesian}, we provide a short review of kernel-based system
identification. In Section~\ref{sec:Estimation}, we propose the
initial-conditions estimation strategies and the related system identification
algorithms. In Section~\ref{sec:experiments}, we
show the results of numerical experiments. In these experiments, the discussed
method are compared with standard techniques used to deal with unknown
initial conditions. In Section~\ref{sec:Discussion}, we summarize the work and
conclude the paper.

\section{Problem formulation}\label{sec:Problem}
We consider the output error model of the form
\begin{equation}\label{eq:system}
 y_t = \sum_{k=0}^\infty g_k u_{t-k} + v_t,
\end{equation}
where ${\{g_t\}}_{t=0}^{+\infty}$ is the impulse response of a linear
time-invariant system. For notational convenience, we assume there are no
delays in the system ($g_0 \neq 0$). We approximate $g$ by considering its first
$n$ samples $\{{g_t\}}_{t=0}^{n-1}$, where $n$ is chosen large enough to capture
the system dynamics. The system is driven by the input $u_t$ and the
measurements of the output $y_t$ are corrupted by the process $v_t$, which is
zero-mean white Gaussian noise with variance $\sigma^2$.

Given a set of $N$ measurements, denoted by ${\{u_t\}}_{t=0}^{N-1}$,
${\{y_t\}}_{t=0}^{N-1}$, we are interested in estimating the first $n$ samples of
the impulse response ${\{g_t\}}_{t=0}^{n-1}$. To this end, we formulate this
system identification problem as the linear regression problem
\begin{equation}\label{eq:system_2}
y = Ug + v \,,
\end{equation}
where we have introduced the following vector/matrix notation
$$
y := \begin{bmatrix} y_0 \\ \vdots \\ y_{N-1} \end{bmatrix} \,,\, g :=
\begin{bmatrix} g_0 \\ \vdots \\ g_{n-1} \end{bmatrix} ,\, v := \begin{bmatrix}
v_0 \\ \vdots \\ v_{N-1} \end{bmatrix},
$$
\begin{equation}\label{eq:matrix_U}
  U = \matrix{u_0 & u_{-1} & u_{-2} & & u_{-n+1}\\
    u_1 & u_0 & u_{-1} &  & u_{-n+2} \\
    u_2 & u_1 & u_{0} &\cdots& u_{-n+3} \\
      \vdots & \vdots  &\vdots &&\vdots\\
      u_{N-1} & u_{N-2}& u_{N-3}&&u_{N-n}
    }.
\end{equation}
The matrix $U$ contains the samples $u_{-1}, \dots, u_{-n+1}$, that we call
\emph{initial conditions}, that are unavailable. Common ways to overcome this
problem are, for instance
\begin{itemize}
\item Discard the first $n-1$ collected samples of $y$. However, if $N$ is not
  much larger than $n$, (e.g., if $n\sim 100$ and $N\sim 200$), there is a
  considerable loss of information.
\item Assume that the system is at rest before the experiment is performed,
  (i.e.  $u_{-1}$, $\ldots$, $u_{-n+1} = 0$). This assumption might be too
  restrictive or unrealistic.
\end{itemize}

In this paper, our aim is to study how to exploit the available information to
estimate the initial conditions, in order to improve the identification
performance.  Specifically, we will present three estimators that make different use
of the available information.

\section{Kernel-based system identification}\label{sec:Bayesian}
In this section we briefly review the kernel-based approach introduced
in~\cite{pillonetto2010new},~\cite{pillonetto2011prediction}. Exploiting the Bayesian interpretation of
kernel-based methods~\cite{wahba1990spline}, we model the unknown impulse response as a
Gaussian random process, namely
\begin{equation}\label{eq:prior_g}
    g \sim \N(0,\lambda K_{\beta}) \,.
\end{equation}
We parameterize the covariance matrix $K_\beta$ (the \emph{kernel}) with the
hyperparameter $\beta$.
The structure of the kernel determines the properties of the realizations of~\eqref{eq:prior_g}; its choice is therefore of paramount importance. An
effective kernel for system-identification purposes is the \emph{stable spline
kernel}~\cite{pillonetto2010new},~\cite{pillonetto2011kernel}. In particular, in this paper we use
the \emph{first-order stable spline kernel} (or \emph{TC kernel}
in~\cite{chen2012estimation}), that is defined as
\begin{equation}\label{eq:ssk1}
  {\{K_\beta\}}_{i,j} := \beta^{ \max(i,j)} \,,
\end{equation}
where $\beta$ is a scalar in the interval $[0,\,1)$. The role of %chktex 9
this hyperparameter is to regulate the velocity of the exponential decay of
the impulse responses drawn from the kernel.  The hyperparameter $\lambda \geq
0$ is a scaling factor that regulates the amplitude of the realizations of~\eqref{eq:prior_g}.

We collect the hyperparameters into the vector
\begin{equation}\label{eq:vector_rho}
\rho := \begin{bmatrix} \lambda & \beta \end{bmatrix}
\end{equation}
and introduce the following notation:
$$
u := \begin{bmatrix} u_- \\ u_+  \end{bmatrix} \quad
u_- := \begin{bmatrix} u_{-n+1} \\ \vdots \\ u_{-1} \end{bmatrix} \quad
u_+ := \begin{bmatrix} u_{0} \\ \vdots \\ u_{N-1} \end{bmatrix} \,,
$$
where $u_-$ contains the unknown initial conditions.
Since we have assumed a Gaussian distribution for the noise, the joint
description of $y$ and $g$ is Gaussian, parameterized by $u_-$ and
$\rho$. Therefore, we can write
\begin{equation}\label{eq:joint_Gaussian}
p\left(\begin{bmatrix} y \\ g \end{bmatrix};\,\rho,\,u_- \right) \sim \mathcal N
\left( \begin{bmatrix} 0\\0 \end{bmatrix} , \begin{bmatrix} \Sigma_y &
\Sigma_{yg} \\ \Sigma_{gy} &  \lambda K_\beta \end{bmatrix} \right)\,,
\end{equation}
where $\Sigma_{yg} = \Sigma_{gy}^T =   \lambda U K_\beta$ and $\Sigma_y =
\lambda  U K_\beta U^T + \sigma^2I$.
It follows that the posterior distribution of $g$ given $y$  is Gaussian, namely
\begin{equation}\label{eq:pg}
p(g|y;\,\rho,\,u_-) = \mathcal N \left(\hat g,\,\Sigma_{g|y} \right) \,,
\end{equation}
where
\begin{equation}\label{eq:CandP}
  \Sigma_{g|y} = {\left( \frac{U^T U}{\sigma^2} +  {(\lambda K_\beta)}^{-1}
\right)}^{-1}\hspace{-0.4em},\qquad \hat g = \Sigma_{g|y} \frac{U^T}{\sigma^2} y  \,.
\end{equation}
Equation~\eqref{eq:pg} implies that the minimum variance estimator  of $g$ (in the Bayesian
sense, see~\cite{anderson2012optimal}) is
\begin{equation}\label{eq:Bayesest_gaussian}
\hat g = \mathbb E [g|y;\,\,\rho,\,u_-]\,.
\end{equation}
The estimate $\hat g$ depends on the hyperparameter vector $\rho$ and the
initial conditions. These quantities need to be estimated from data.
In the next section we focus our attention to the estimation of the kernel
hyperparameters and the initial conditions, describing different strategies to
obtain these quantities.
\begin{remark}
The estimator~\eqref{eq:Bayesest_gaussian} depends also on the noise variance
$\sigma^2$. In this work, we assume that this parameter is known. It can for
instance be estimated by fitting a least-squares estimate of the system $g$ and
then computing the sample variance of the residuals.\qed\
\end{remark}

\section{Estimation of initial conditions and hyperparameters}\label{sec:Estimation}
In most works on kernel-based system identification (see e.g.~\cite{pillonetto2014kernel} for a survey), the authors adopt an empirical-Bayes
approach to estimate the hyperparameters that define the kernel. This amounts to
maximizing the marginal likelihood (ML) of the output, found integrating $g$ out
of~\eqref{eq:joint_Gaussian}.

In the standard case, that is when $u_-$ is assumed to be known, the ML
estimator of the hyperparameters corresponds to
\begin{equation}\label{eq:ML}
  \hat \rho = \arg\max_\rho p(y;\rho,u_-)\,.
\end{equation}
We start from~\eqref{eq:ML} to design new estimators for the initial conditions and the kernel hyperparameters.

\subsection{Model-less estimate}\label{ssec:Naive}
The most straightforward generalization of~\eqref{eq:ML} is to include the
initial conditions among the ML parameters. The initial conditions become
 unknown quantities that parameterize the impulse response estimator.
The ML criterion then becomes
\begin{equation}\label{eq:ML_naive}
\hat{\rho},\, \hat{u}_- = \arg \max_{\rho,\,u_-} p(y\,;\,\rho,\,u_-)\,,
\end{equation}
where the maximization is carried out over the unknown initial conditions as
well. This problem is nonconvex and possibly high dimensional, as the number of
initial conditions to be estimated is equal to the number of impulse response
samples. To overcome this difficulty, we devise a strategy based on the
expectation-maximization method that yields a solution to~\eqref{eq:ML_naive} by
iterating simple updates.  To this end, we suppose we have any estimate of the
unknown quantities $\hat\rho^{(k)}$ and $\hat u_{-}^{(k)}$, and we calculate the
current estimate of the impulse response, as well as its variance, from~\eqref{eq:CandP}.  Define the matrix
\begin{equation}
  \hat S^{(k)} = \hat \Sigma_{g|y}^{(k)} + \hat g^{(k)}\hat g^{(k)\,T},
\end{equation}
which is the second moment of the current estimated impulse response.
We introduce the \emph{discrete-derivator} matrix
% \begin{equation}\label{eq:matrix_T}
% \Delta := \begin{bmatrix}
%       1 & -1      &     & 0\\
%           & 1   &   \ddots  & \\
%         &   & \ddots  &-1 \\
%       0 &     &     & 1
%     \end{bmatrix} \,,
% \end{equation}
\begin{equation}\label{eq:matrix_T}
  \Delta_{ij} = \delta_{i,j} - \delta_{i,j+1},
\end{equation}
and we calculate the second moment of the derivative of the estimated impulse response at iteration
$k$, $\Delta \hat g^{(k)}$, given by
\begin{equation}\label{eq:D}
  \hat D^{(k)}  :=  \Delta \hat S^{(k)}\Delta^T\,.
\end{equation}

The Toeplitz matrix of the input samples $U$ can be split in two
parts, namely $U = U_+ + U_-$, where $U_+$ is fully determined by the available samples,
and $U_-$ is composed of the unknown initial conditions. Define the
matrix $\R\in \mathbb{R}^{Nn\times N}$ that satisfies the relation
$\R u_- = \mathrm{vec}(U_-)$; and call $\hat G^{(k)}$ the Toeplitz matrix of the
estimated impulse response at the $k$th iteration. Furthermore, define
\begin{align}\label{eq:matrix_A_B}
  \hat A^{(k)} \! & := \!\R^T\Big[ \hat S^{(k)} \kron I_N\Big]\R,\\
  \hat b^{(k)\,T}\! &:= \!
  \mathrm{vec}{(U_+)}^T\Big[ \hat S^{(k)} \kron I_N\Big]\R-y^T\hat G^{(k)} \!.\nonumber
\end{align}
With all the definitions in place, we can state the theorem that provides us
with the iterative update of the estimates that solves~\eqref{eq:ML_naive}.
\begin{theorem}\label{thm:Naive}
  Consider the hyperparameter estimator~\eqref{eq:ML_naive}.
  Starting from any initial guess of the initial conditions and the hyperparameters,
  compute
  \begin{align}\label{eq:uhat_Naive}
    \hat u^{(k+1)} &= {\big(\hat A^{(k)}\big)}^{-1}\hat b^{(k)},\\
    \hat \beta^{(k+1)} &= \arg \min_{\beta \in [0,1)} Q(\beta) %chktex 9
    \,,\label{eq:min_beta}\\
    \hat \lambda^{(k+1)} &= \frac{1}{n} \sum_{i=1}^n  \hat d_i^{(k)} w_{\hat \beta^{(k+1)},i}\;,\label{eq:min_lambda}
  \end{align}
  with $\hat A^{(k)}$ and $\hat b^{(k)}$ defined in~\eqref{eq:matrix_A_B}, and
    \begin{align}\label{eq:Q_beta}
      Q(\beta) & := n \log f(\beta) +  \frac{n(n-1)}{2} \log \beta -
      \log(1-\beta) \,,\\
  f(\beta) &:= \sum_{i=1}^{n-1} \hat d_i^{(k)} \beta^{-i} + \hat d_n^{(k)}(1-\beta)\beta^{1-n} \,;\label{eq:function_f}
\end{align}
    where $\hat d_i^{(k)}$ is the $i$th diagonal element of~\eqref{eq:D}, and
    $w_{\hat \beta^{(k+1)},i}$ is the $i$th element of
      \begin{equation}\label{eq:vector_w}
      w_\beta  := \frac{1}{\beta - \beta^2} \begin{bmatrix} 1 & \frac{1}{\beta}
      & \cdots & \frac{1}{\beta^{n-2}} & \frac{1-\beta} {\beta^{n-1}}
    \end{bmatrix},
      \end{equation}
     when $\beta = \hat \beta^{(k+1)}$.
  Let $\hat\rho^{(k+1)}=[\hat\lambda^{(k+1)},\hat\beta^{(k+1)}]$; then
  the sequences ${\{\hat u^{(k)}\}}_{k=0}^{\infty}$ and
  ${\{\hat \rho^{(k)}\}}_{k=0}^{\infty}$ converge to a maximum of~\eqref{eq:ML_naive}.
\end{theorem}
\begin{proof}
  See Appendix.
\end{proof}
\begin{remark}
The EM method does not guarantee convergence of the sequences to a global
maximum (see~\cite{mclachlan2007algorithm} and~\cite{tseng2004analysis}). However,
experiments (see Section~\ref{sec:experiments}) have show that, for this particular problem, the EM
method converges to the global maximum independently of how it is initialized.
\end{remark}

Thus, we can use Theorem~\ref{thm:Naive} to find a maximum of the marginal
likelihood of the hyperparameters and the initial conditions, and then use these
parameters to solve the impulse response estimation problem with~\eqref{eq:Bayesest_gaussian}, where we use the limits of the sequences of
estimates.

\subsection{Conditional mean estimate}\label{ssec:CondMean}
The model-less estimator presented in Section~\ref{ssec:Naive} estimates the initial
conditions using only information present in the system output $y$. It does not rely on
any model of the input signal $u$. To show how an available model can be used
to estimate the missing initial conditions, we introduce the following
assumption.
\begin{assumption}\label{assumption1}
The input $u_t$ is a realization of a stationary Gaussian process with zero-mean
and known rational spectrum. Equivalently, $u_t$ is a realization on an ARMA
process with known coefficients.\qed{}
\end{assumption}
Assumption~\ref{assumption1} implies that $u_t$ can be expressed as the output
of a difference equation driven by white Gaussian noise with unit
variance~\cite{papoulis2002probability}, namely
\begin{equation}\label{eq:ARMA}
u_t + d_1 u_{t-1} + \cdots + d_p u_{t-p} = c_0 e_t + \cdots + c_p e_{t-p} \,,
\end{equation}
where $e_t \sim \mathcal{N}(0,\,1)$.
Since, using Assumption~\ref{assumption1}, we can construct the probability
density of the input process, a possible approach to solve~\eqref{eq:ML} is to
estimate the missing initial conditions from the input process.
To this end, consider~\eqref{eq:ARMA}. If we define
the matrices $D$ as the toeplitz matrix of the coefficients
$0,d_1,d_2,\dots$, and  $C$ as the toeplitz matrix of the coefficients
$c_1,c_2,\dots$,
% $$
% D := \begin{bmatrix}
% 0     &     & \ldots & 0 \\
% d_1   & 0   & \ldots & 0 \\
% d_2   & d_1 & 0 &     \vdots \\
% \vdots  & \ddots& \ddots & \ddots
% \end{bmatrix} \,,\,
% C := \begin{bmatrix}
% c_0     &     & \ldots & 0 \\
% c_1   & c_0   & \ldots & 0 \\
% c_2   & c_1 & c_0 &     \vdots \\
% \vdots  & \ddots& \ddots & \ddots
% \end{bmatrix} \,;
% $$
then we can write
\begin{equation}
  u = -Du + Ce \,, \quad e:= \begin{bmatrix} e_{-n+1} & \cdots & e_N
  \end{bmatrix}^T \,,
\end{equation}
so that $p(u) \sim \mathcal{N}(0,\,\Sigma_u)$, with
\begin{equation}
 \label{eq:Sigma_u}
 \Sigma_u = {(I+D)}^{-1}CC^T{(I+D)}^{-T}.
\end{equation}
We thus have a joint probabilistic description of the initial conditions $u_-$ and
the available input samples $u_+$. We can partition $\Sigma_u$ into four blocks
according to the sizes of $u_-$ and $u_+$
$$
\Sigma_u = \begin{bmatrix}
\Sigma_- & \Sigma_{-+} \\
\Sigma_{+-} & \Sigma_+
\end{bmatrix} \,.
$$
It follows (see~\cite{anderson2012optimal}) that the posterior
distribution of the unavailable data is $p(u_-|u_+) = 
\mathcal{N}(u_{-|+},\,\Sigma_{-|+})$, where
\begin{equation}\label{eq:conditional_u}
u_{-|+} = \Sigma_{-+}\Sigma_+^{-1} u_+ \,,\quad \Sigma_{-|+} = \Sigma_- -
\Sigma_{-+}\Sigma_+^{-1}\Sigma_{+-}\,.
\end{equation}
So we can  find the minimum variance estimate of $u_-$ as the conditional mean
$u_{-|+}$, namely $\hat u_- = u_{-|+}$.

Having an estimate of the initial conditions, we need to find the
hyperparameters that define the kernel. In this case, empirical Bayes amounts
to solving the ML problem
\begin{equation}\label{eq:ML_CondMean}
\hat{\rho} = \arg \max_{\rho} p(y\,;\,\rho,\,\hat u_-)\,,
\end{equation}
where the unknown initial conditions have been replaced by their conditional
mean. The following theorem states how to solve the maximization using the EM
method.
\begin{theorem}\label{thm:CondMean}
  Consider the hyperparameter estimator~\eqref{eq:ML_CondMean}. Starting from an
  initial guess of the hyperparameters, compute
  \begin{align}
    \hat \beta^{(k+1)} &= \arg \min_{\beta \in [0,1)} Q(\beta) %chktex 9
    \,,\label{eq:CondMean_beta}\\
    \hat \lambda^{(k+1)} &= \frac{1}{n} \sum_{i=1}^n  \hat d_i^{(k)}
    w_{\hat \beta^{(k+1)},i}\label{eq:CondMean_lambda}\,,
  \end{align}
  with $Q(\beta)$, $\hat d_i^{(k)}$, and $w_{\hat \beta^{(k+1)},i}$ defined in
  Theorem~\ref{thm:Naive}.
  Let $\hat\rho^{(k+1)}=[\hat\lambda^{(k+1)},\hat\beta^{(k+1)}]$,
  then the sequence
  ${\{\hat \rho^{(k)}\}}_{k=0}^{\infty}$ converges to a maximum of~\eqref{eq:ML_CondMean}.
\end{theorem}
\begin{proof}
  See Appendix.
\end{proof}
\begin{remark}
  The updates~\eqref{eq:CondMean_beta} and~\eqref{eq:CondMean_lambda} require the
  evaluation of~\eqref{eq:D} at each iteration. In this case the estimate
  $\hat g^{(k)}$ of the impulse response is given by~\eqref{eq:Bayesest_gaussian},
  where $u_-$ is replaced by its conditional mean.\qed{}
\end{remark}

\subsection{Joint input-output estimate}\label{ssec:Joint}
The conditional mean estimator presented in Section~\ref{ssec:CondMean} exploits
the structure of the input to estimate the missing samples. The model-less estimator,
instead, uses information contained in the output samples. In this section we
show how to merge these two information sources, defining a joint input-output
estimator of the initial conditions.

We use Assumption~\ref{assumption1} to account for the statistical
properties of $u_-$. We propose the following mixed MAP--ML estimator %chktex 8
\begin{equation}\label{eq:ML_Joint}
\hat{\rho},\, \hat{u}_- = \arg \max_{\rho,\,u_-}
p(y\,|\,u_-,\,u_+;\,\rho)p(u_-|\,u_+)\,,
\end{equation}
where we have highlighted the dependence on the known input sequence $u_+$. A
key role is played by the term $p(u_-|\,u_+)$: it acts as a prior distribution
for the unknown values of $u_-$ and puts weight on the values that better agree
with the observed data $u_+$.

Even in this case, the solution can be found with an iterative procedure based
on the EM method.
\begin{theorem}\label{thm:Joint}
  Consider the hyperparameter estimator~\eqref{eq:ML_Joint}.  Starting from an
  initial guess of the initial conditions and the hyperparameters,
  compute
  \begin{align} 
    \hat u^{(k+1)} &= {\left(\!\frac{{\big(\hat A^{(k)}\big)}^{-1}}{\sigma^2}\! +
  \!    \Sigma_{-|+}^{-1} \!\right)}^{\!\!\!-1}
    \!\!\!\left(\!\frac{\hat b^{(k)}}{\sigma^2\!}+\! \Sigma_{-|+}^{-1}\!
    u_{-|+}\!\right)\,,\label{eq:uhat_Joint}\\
    \hat \beta^{(k+1)} &= \arg \min_{\beta \in [0,1)} Q(\beta) % chktex 9
    \,,\label{eq:Joint_beta}\\
    \hat \lambda^{(k+1)} &= \frac{1}{n} \sum_{i=1}^n  \hat d_i^{(k)}
    w_{\hat \beta^{(k+1)},i}\label{eq:Joint_lambda}\,,
  \end{align}
  with $\hat A^{(k)}$ and $ \hat b^{(k)}$ from~\eqref{eq:matrix_A_B}; and
  with $Q(\beta)$, $\hat d_i^{(k)}$, and $w_{\hat \beta^{(k+1)},i}$ defined in
  Theorem~\ref{thm:Naive}.
  Let $\hat\rho^{(k+1)}=[\hat\lambda^{(k+1)},\hat\beta^{(k+1)}]$;
  then, the sequences ${\{\hat u^{(k)}\}}_{k=0}^{\infty}$ and
   ${\{\hat \rho^{(k)}\}}_{k=0}^{\infty}$ converge to a maximum of~\eqref{eq:ML_Joint}.
\end{theorem}
\begin{proof}
  See Appendix.
\end{proof}
\begin{remark}
  This estimator incorporates the prior information about the initial conditions
  using the mean $u_{-|+}$ and the covariance matrix $\Sigma_{-|+}$.
  If we suppose that we can manipulate $\Sigma_{-|+}$, we can see this estimator as a
  more general estimator, that contains the model-less and conditional mean as
  limit cases. In fact, setting $\Sigma_{-|+}=\infty$, we get the model-less
  estimator. Conversely, setting $\Sigma_{-|+}=0$, we obtain the
  conditional-mean estimator, as~\eqref{eq:uhat_Joint} would
  yield a degenerate iteration where all the updates are $\hat u^{(k+1)} =
  u_{-|+}$. We point however out that the model-less estimator does not rely on any
  assumption on the input model, whereas the joint input-output estimator requires
  that the input is a Gaussian process with known pdf.
\end{remark}

\section{Numerical experiments}\label{sec:experiments}
\subsection{Experiment setup}
To compare the proposed methods, we perform five numerical experiments, each one
consisting of 200 Monte Carlo simulations, with sample sizes 150, 200, 250,
300, and 400. At each Monte Carlo run, we generate
a dynamic system of order 40. The system is
such that the zeros are constrained within the circle of radius 0.99 on the
complex plane, while the magnitude of the poles is no larger than 0.95. The
impulse response length is 100 samples. The input is obtained by filtering white
noise with unit variance through a 8-th order ARMA filter of the form~\eqref{eq:ARMA}. The coefficients of the filter are randomly chosen at each
Monte Carlo run, and they are such that the poles of the filter are constrained
within the circular region of radii 0.8 and 0.95 in the complex plane.

Random trajectories of input and noise are generated at each run. In
particular, the noise variance is such that the ratio between the variance of
the noiseless output of the system and the noise variance is equal to 20.

The following estimators are compared during the experiments.
\begin{itemize}
\item \emph{KB-IC-Zeros}: This method does not attempt any estimation of the
  initial conditions. It sets their value to 0 (that, when assumption~\ref{assumption1} holds, corresponds to the a-priori mean of the vector
$u_-$). The kernel hyperparameters are obtained solving~\eqref{eq:ML}, with $u_-
= 0$.
\item \emph{KB-Trunc}: This method also avoids the estimation of
  the initial conditions by discarding the first $n-1$ output samples, which
  depend on the unknown vector $u_-$. The hyperparameters are obtained solving~\eqref{eq:ML}, using the truncated data.
\item \emph{KB-IC-ModLess}: This is the model-less kernel-based estimator presented in
  Section~\ref{ssec:Naive}.
\item \emph{KB-IC-Mean}: This is the conditional mean kernel-based estimator presented in
  Section~\ref{ssec:CondMean}.
\item \emph{KB-IC-Joint}: This is the joint input-output kernel-based estimator presented in
  Section~\ref{ssec:Joint}.
\item \emph{KB-IC-Oracle}: This estimator has access to the vector $u_-$, and
  estimates the kernel hyperparameters using~\eqref{eq:ML}.
  \end{itemize}

The performances of the estimators are evaluated by means of the fitting score, computed as
\begin{equation}
  FIT_i  = 100\left(1-\frac{\|g_i - \hat g_i \|_2}{\|g_i - \bar g_i\|_2}\right) \,,
\end{equation}
where $g_i$ is the impulse response generated at the $i$-th run, $\bar g_i$ its
mean and $\hat g_i$ the estimate computed by the tested methods.
\subsection{Results}

\begin{table}
  \centering
  \begin{tabular}{cccccc}
    \toprule
  $N$  & 150 & 200 & 250 & 300 & 400 \\
  \midrule
  KB-IC-Zeros   & 51.698 & 54.856 & 61.151 & 61.380 & 63.186 \\
  KB-Trunc      & 42.010 & 51.038 & 59.400 & 61.085 & 62.963 \\
  KB-IC-ModLess   & 54.017 & 55.793 & 61.687 & 63.074 & 63.466 \\
  KB-IC-Mean    & 54.146 & 56.003 & 62.061 & 63.715 & 64.187 \\
  KB-IC-Joint   & 55.695 & 57.133 & 62.776 & 64.310 & 64.457 \\
  KB-IC-Oracle  & 57.317 & 57.902 & 63.781 & 64.893 & 64.959 \\
  \bottomrule
  \end{tabular}
  \caption{Table of experimental results. Shown is the average fit in percent over the
  different experiments.}\label{tab:exp_results2}
\end{table}

Table~\ref{tab:exp_results2} shows the average fit (in percent) of the impulse
response over the Monte Carlo experiments. We can see that, for short data
sequences  the amount of information discarded by the estimator KB-Trunc makes
its performance degrade with respect to the other estimators. The estimator
KB-IC-Zeros performs better, however suffers from the effects of the wrong
assumption that the system was at rest before the experiment was performed.
From these results, we see that the estimation of the initial conditions has a
positive effect on the accuracy of the estimated impulse response.  For larger
data records  the performances of the estimator
KB-IC-Mean and of the estimator KB-IC-ModLess improve, as more samples become
available.

When the available data becomes larger, all the methods perform
well, with fits that are in the neighborhood of the fit of the oracle.

The positive performance of KB-IC-Mean indicates that the predictability of the
input can be exploited to improve estimates, and that model-based approaches to
initial condition estimation outperforms model-less estimation methods (if the input
model is known). The further improvement of KB-IC-Joint indicates that the
output measurements can be used to obtain additional information about the unobserved
initial conditions, information that is not contained in the input process itself.

\section{Discussion}\label{sec:Discussion}
We have proposed three new methods for estimating the initial conditions of a system
when using kernel-based methods. Assuming that the input is a stationary ARMA
process with known spectrum, we have designed mixed MAP--ML criteria %chktex 8
which aim at jointly estimating the hyperparameters of the kernel and the
initial conditions of the systems. To solve the related optimization problems,
we have proposed a novel EM-based iterative scheme. The scheme consists in a
sequence of simple update rules, given by unconstrained quadratic problems or
scalar optimization problems. Numerical experiments have shown that the proposed
methods outperform other standard methods, such as truncation or zero initial
conditions.

The methods presented here estimate $n-1$ initial conditions (where $n$ is the
length of the FIR approximating the true system), since no information on the
order of the system is given. Assuming that the system order is known and equal
to say, $p$, the number of initial conditions to be estimated would boil down to
$p$. However, there would also be $p$ unknown transient responses which need to
be identified. These transients would be characterized by impulse responses with
the same poles as the overall system impulse response, but with different zeros.
How to design a kernel correlating these transient responses with the system
impulse response is still an open problem.

\section*{Appendix: Proofs}
\subsection{Theorem~\ref{thm:Naive}}
Consider the ML criterion~\eqref{eq:ML_naive}. To apply the EM method, we
consider the complete log-likelihood
\begin{align}\label{eq:proof_naive_loglik}
  L(y,g) &= -\frac{1}{2\sigma^2}\| y - Ug\|^2 - \frac{N}{\sigma^2}\log \sigma^2
  \nonumber \\
  &-\frac{1}{2}g^T{\big( \lambda K_{\beta} \big)}^{-1}g -
  \frac{1}{2}\log\det \big(\lambda K_{\beta}\big).\nonumber
\end{align}
where we have introduced $g$ as a latent variable.
Suppose that we have computed the estimates $\hat \rho^{(k)}$ of the
hyperparameters and $\hat u_-^{(k)}$ of the initial conditions. We define the
function
\begin{equation}\label{eq:E_step}
  Q(\rho,u_-;\,\hat \rho^{(k)},\hat u_-^{(k)}) := \E \left[
  L(y,\,g)\right]\,,
\end{equation}
where the expectation is taken with respect to the conditional density
$p(g|y;\,\hat \rho^{(k)},\hat u_-^{(k)})$, defined in~\eqref{eq:pg}.
We obtain (neglecting terms independent from the optimization variables)
\begin{equation}\label{eq:Q1Q2}
  Q(\rho,u_-;\,\hat \rho^{(k)}, \hat u_-^{(k)}) \!= \!Q_1(
  u_-;\hat \rho^{(k)},\hat u_-^{(k)} ) + Q_2( \rho;\hat \rho^{(k)},\hat u_-^{(k)} ),
\end{equation}
where 
\begin{align}
  Q_1(u_-;\hat \rho^{(k)},\hat u_-^{(k)} ) &= \nonumber\\
  &\hspace{-7em}-\frac{1}{2\sigma^2}\Big(y^T U_-\hat g +\Tr \Big\{
  \big(U_{-}^T U_- -2 U_-^T U_{+}\big)\,\hat S^{(k)}\Big\}\Big )\,,\label{eq:Q_1proof}\\
  \!\!\!Q_2(\rho;\hat \rho^{(k)},\hat u_-^{(k)}) &=- \frac{1}{2}\Tr\Big\{\!(\!\lambda
  K_\beta)\!^{-1}\,\hat S^{(k)} \!\Big\} -
  \frac{1}{2}\log\det\big(\lambda K_\beta),\label{eq:Q_2proof} 
\end{align}
and $\hat S^{(k)} = \hat\Sigma_{g|y}+\hat g\,\hat g^T$.
Define new parameters from
\begin{equation}\label{eq:M_step}
  \hat \rho^{(k+1)},\hat u_-^{(k+1)}= \arg\max_{\rho,u_-}
  \mathcal{Q}(\rho,u_-;\,\hat \rho^{(k)},\hat u_-^{(k)}) \,.
\end{equation}
By iterating between~\eqref{eq:E_step} and~\eqref{eq:M_step} we obtain a
sequence  of estimates that converges to a maximum of~\eqref{eq:ML_naive} (see
e.g.,~\cite{mclachlan2007algorithm} for details).
Using~\eqref{eq:Q1Q2},~\eqref{eq:M_step} splits in the two maximization problems
\begin{align}
  \hat u_-^{(k+1)} &= \arg\max_{u_-}
  Q_1(u_-;\,\hat \rho^{(k)},u_-^{(k)})\,\label{eq:Q_1},\\
  \hat\rho^{(k+1)} &= \arg\max_{\rho} Q_2(\rho;\,\hat \rho^{(k)}, \hat u_-^{(k)}) \,.\label{eq:Q_2}
\end{align}

Consider the maximization of~\eqref{eq:Q_1proof} with respect to $u_-$. Using
the matrix $\R$ we have
\begin{align}
  \Tr\Big\{ U_-^T U_-\hat S^{(k)}\} &= u_-^T \R^T\Big[ \hat S^{(k)} \kron I_N\Big]\R u_- \,,\\
  \Tr\Big\{ U_-^T U_+\hat S^{(k)}\Big\}&= {\mathrm{vec}(U_+)}^T \Big[ \hat S^{(k)} \kron I_N\Big]\R
  u_- \,,
\end{align}
where $\kron$ denotes the Kronecker product.
We now collect $u_-$, and obtain
\begin{equation}\label{eq:Q1_quadratic}
  \quad Q_1(u_-;\hat \rho^{(k)},\hat u_-^{(k)} ) = -\frac{1}{2 \sigma^2}u_-^T
  \hat A^{(k)}u_- + \frac{1}{\sigma^2}u_-^T\hat b^{(k)}
\end{equation}
with $\hat A^{(k)}$ and $ \hat b^{(k)}$ defined in~\eqref{eq:matrix_A_B}.
Hence,~\eqref{eq:Q_1} is an unconstrained quadratic optimization, whose solution is
given by~\eqref{eq:uhat_Naive}.

Consider now the maximization of~\eqref{eq:Q_2proof} with respect to $\rho$.
We can calculate the derivative of $Q_2$ with respect to $\lambda$, obtaining
\begin{equation}\label{eq:partial_lambda}
  \frac{\partial Q_2}{\partial \lambda} =
  -\frac{1}{2\lambda^2}\Tr\Big\{K_\beta^{-1} \hat S^{(k)} \Big\} + \frac{n}{2\lambda};
\end{equation}
which is equal to zero for
\begin{equation}\label{eq:lambda_opt}
  \lambda^* = \frac{1}{n} \left(\Tr\Big\{K_\beta^{-1} \hat S^{(k)} \Big\}\right) \,.
\end{equation}
We thus have an expression of the optimal value of $\lambda$ as a function of $\beta$. If
we insert this value in $Q_2$, we obtain
\begin{align}
  Q_2([\lambda^*\,  \beta];\,\hat \rho^{(k)},\hat u_-^{(k)}) & =
  \nonumber\\
  &\hspace{-9em}-\frac{1}{2}\log\left(\Tr\Big\{K_\beta^{-1} \hat S^{(k)} \Big\}\right) -\frac{1}{2}\log \det
  K_\beta + k_1,
\end{align}
where $k_1$ is constant.
We now rewrite the first order stable spline kernel using the factorization
(see~\cite{carli2014maximum})
\begin{equation}\label{eq:factorization_peopeo}
  K_\beta = \Delta^{-1} W_\beta \Delta^{-T} \,,
\end{equation}
where $\Delta$ is defined in~\eqref{eq:matrix_T} and
\begin{equation}
  W_\beta  := (\beta - \beta^2) \diag\left\{1,\,\beta,\,\ldots,\,\beta^{n-2},\,\frac{\beta^{n-1}}{1-\beta}\right\} \,.
\end{equation}
From~\eqref{eq:D}, we find
\begin{align}\label{eq:Q_after_rewrite}
  Q_2([\lambda^*\,  \beta];\,\hat \rho^{(k)},\hat u_-^{(k)}) & =\nonumber\\
  n \log \left( \sum_{i=1}^{n} \hat d_{i}^{(k)} w_{\beta,ii}^{-1}\right)
  & +  \sum_{i=1}^{n} \log  w_{\beta,ii} + k_2\,,
\end{align}
where $\hat d_{i}^{(k)}$ and $w_{\beta,ii}$ are the $i$-th diagonal elements of
$\hat d^{(k)}$ and of $W_{\beta}$ respectively, and $k_2$ is a constant.
If we define the function~\eqref{eq:function_f}, we can
rewrite~\eqref{eq:Q_after_rewrite} as~\eqref{eq:Q_beta}.
so that we obtain~\eqref{eq:Q_beta} and~\eqref{eq:function_f}. Using a similar
reasoning, we can rewrite~\eqref{eq:lambda_opt} as~\eqref{eq:min_lambda}.
\qed{}

\subsection*{Theorem~\ref{thm:CondMean}}
Consider the ML criterion~\eqref{eq:ML_CondMean}. The proof follows the same
arguments as the proof of Theorem~\ref{thm:Naive}, with the optimization
carried out on $\rho$ only and with $u_-=u_{-|+}$.\qed{}
\subsection*{Theorem~\ref{thm:Joint}}
Consider the ML criterion~\eqref{eq:ML_Joint}. Consider the complete data
log-likelihood
\begin{equation}
  L_2(y,g) := \log p(y,\,g,\,u_-;\,\rho,)\,,\\
\end{equation}
Given any estimates $\hat \rho^{(k)}$ and $\hat u_-^{(k)}$, take the expectation
with respect to $p(g|y;\hat \rho^{(k)},\hat u_-^{(k)})$. We obtain
\begin{align}
\E\big[L_2(y,g)\big]& =  Q_1( u_-;\hat \rho^{(k)},\hat u_-^{(k)} ) +
  Q_2( \rho;\hat \rho^{(k)},\hat u_-^{(k)} )\nonumber\\
  &-\frac{1}{2}{\big(u_-\!\!-  u_{-|+}\big)}^T\Sigma_{-|+}^{-1}\big(u_-\!\!-
u_{-|+}\big)\,,
\end{align}
with $Q_1$ and $Q_2$ defined in~\eqref{eq:Q_1proof} and~\eqref{eq:Q_2proof}.
Collecting the terms in $u_-$ and using~\eqref{eq:Q1_quadratic} we obtain an
unconstrained optimization problem in $u_-$, that gives~\eqref{eq:uhat_Joint}.
The optimization in $\rho$ follows the same procedure as in
Theorem~\ref{thm:Naive} and gives~\eqref{eq:Joint_beta} and~\eqref{eq:Joint_lambda}.

\small
\printbibliography{}
%===============================================================================%
\end{document}